\documentclass{elsart}
\usepackage{graphicx}
\usepackage{amsmath}
\usepackage{amsfonts}
\usepackage{amssymb}

\begin{document}

\begin{frontmatter}
\title{Role of surface disorder on the magnetic properties and hysteresis of nanoparticles}

\author{\`{O}scar Iglesias}
\ead{oscar@ffn.ub.es}
\ead[url]{http://www.ffn.ub.es/oscar}
and 
\author{Am\'{\i}lcar Labarta} 
\address{Departament de F\'{\i}sica Fonamental, Universitat de Barcelona, Diagonal 647, 08028 Barcelona, Spain}

\begin{abstract}
We present the results of Monte Carlo simulations of a model of a single maghemite ferrimagnetic nanoparticle including radial surface anisotropy distinct from that in the core
with the aim to clarify what is its role on the magnetization processes at low temperatures. 
The low temperature equilibrium states are analized and compared to those of a ferromagnetic particle with the same lattice structure.
We have found that the formation of hedgehog-like structures due to increased surface anisotropy is responsible for a change in the reversal mechanism of the particles.
\vspace{1pc}
\end{abstract}

\end{frontmatter}
\section{Introduction}
The progressive reduction of particle sizes with technological application down to the nanometric range results in magnetic behaviours that are predominantly dominated by the particle surface spins. The magnetic behaviour of the particle surface differs from that corresponding to the core because of the lower coordination number and the existence of broken exchange bonds. Among other effects, this results in increased anisotropy compared to bulk values due to the surface anisotropy with uniaxial normal direction that comes from symmetry breaking at the boundaries of the particle \cite{Batllejpd02}. Moreover, depending on the character of the exchange interaction between spins in the particle and the geometry of their underlying lattice, surface spins may suffer random canting and form a spin-glass state as has been argued in studies of fine particles of different ferrimagnetic oxides \cite{Coeyprl71}.
These systems display anomalous magnetic properties at low temperatures. 
In particular, they display high closure fields with high values of the differential susceptibility and shifts in the hysteresis loops after field cooling \cite{Kodamaprl97,Kodamaprb99,Montseprb99,Martinezprl98} whose origin is still a matter of controversy.
While some of these features have been attributed the existence of dipolar interactions among the particles \cite{Batlleprb97,Dormannjm98,Morupprl94,Jonssonprb00}, there are experimental evidences that finite-size and surface effects are crucial in order to understand this phenomenology.
In this study, we present the results of Monte Carlo (MC) simulations of the magnetic properties of individual nanoparticles which aim at clarifying what is the specific role played by increased radial anisotropy at the surface on the magnetization processes.  

\section{Model}

The model considered is an extension to Heisenberg spins of our previous MC simulations for an Ising spin model of a $\gamma$-Fe$_2$O$_3$ particle \cite{Iglesiasprb01,Iglesiasjap01}. The particles considered are spheres of diameter $D$ (in units of the linear unit cell size $a$).
The simulations have been run for ferrimagnetic particles with maghemite lattice structure and compared with those corresponding to ferromagnetic (FM) particles with the same underlying lattice. 
In order to characterize the specific contribution of the surface spins to the magnetic properties of the particle, we have distinguished two different regions in the particle: the surface formed by the outermost unit cells and an internal core of diameter $D_{Core}= D-2$.

The magnetic ions are modeled by Heisenberg classical spins 
with the following interaction Hamiltonian:
\begin{eqnarray}
\label{Eq1}
{ H}/k_{B}= 
-\sum_{\langle  i,j\rangle}J_{ij} {\vec S}_i \cdot {\vec S}_j   
-\sum_{i= 1}^{N} \vec H\cdot{\vec S_i}
-\sum_{i= 1}^{N}\left[k_C(S_i^z)^2+k_S(\vec{S}_i \cdot \hat n_i)^2 \right]\ , 	
\end{eqnarray}
which is an extension of our previous study with Ising spins \cite{Iglesiasprb01}. 
Here, $\vec S_i$ are three dimensional classical vector spins placed on a discrete lattice and 
$\vec H$ is the magnetic field, that in the following will be given in temperature units as $\vec h= \mu \vec H / k_B$, with $\mu$ the moment of the magnetic ion. 
The $J_{ij}$'s are the nearest neigbour exchange constants, which might vary from site to site depending on the specific material considered, for the case of maghemite, which is a two sublattice (O,T) ferrimagnet, their values are $J_{TT}=-21$ K, $J_{OO}= -8.6$ K, $J_{TO}= -28.1$ K \cite{Iglesiasprb01,Kodamaprb99,Kachkachiepj00}.
The last term in Eq. \ref{Eq1} accounts for the magnetic anisotropy energy, which has been taken as uniaxial along the z-axis for the core spins and along the radial direction for the spins at the surface. $k_C$ and $k_S$ are the core and surface anisotropy constants respectively in temperature units. Although the values of bulk anisotropy constants $K_C$ can be obtained indirectly by magnetic measurements, the surface contribution $K_S$ to the anisotropy energy is more difficult to evaluate. However, for maghemite $K_C\simeq 4.7\times 10^{4}$ erg/cm$^3$ \cite{Krupricka75,Vassilioujap93} and $K_S$ has been estimated  as $K_S\simeq 0.06$ erg/cm$^2$ from M\"ossbauer experiments \cite{Dormannprb96,Gazeaueul97}. Therefore, in the simulations for the maghemite particle, we will vary the core anisotropy values in the range $k_C= 0.01-1$ K and those of the surface anisotropy in the range $k_S= 1-100$ K.

The simulations have been performed using the standard Metropolis algorithm \cite{Landaubook00} for continuous spins. As for the details, let us only mention that, in this case, special care has to be taken in the update algorithm with respect to the case of Ising spins. In order to sample efficiently the configuration space both in the low and high anisotropy limits, we have chosen a sequential combination of three kinds of trial angles in a similar way as Hinzke and Nowak \cite{Hinzkecpc99}. The update algorithm consists of a series of three uniform trial steps at random, one small trial step inside a cone around the initial spin direction and a reflection of the spin direction at the x-y plane to guarantee the correct performance of the simulation at high values of the anisotropy.
\section{Results}

In order to gain insight on the equilibrium properties of particles with surface anisotropy , we have performed a study of the equilibrium configurations in zero magnetic field at low temperatures. For this purpose, the system is initially prepared in a high temperature state with spins randomly oriented and then it is subsequently cooled in constant temperature steps $\delta T=1$ down to zero temperature. At each temperature we measure the thermal averages of the quantities of interest such as the total energy, sublattice magnetizations ($M_O$, $M_T$), surface $M_S$ and core $M_C$ contributions to the total magnetization $M_{Total}$, and also $M_{n}=\sum_{i=1}^N \left| \vec S_i\cdot \hat n_i \right|$,
the sum of projections of the spins onto the local anisotropy axis. 

\begin{figure}[tbp] 
\centering 
\includegraphics[width=0.45\textwidth]{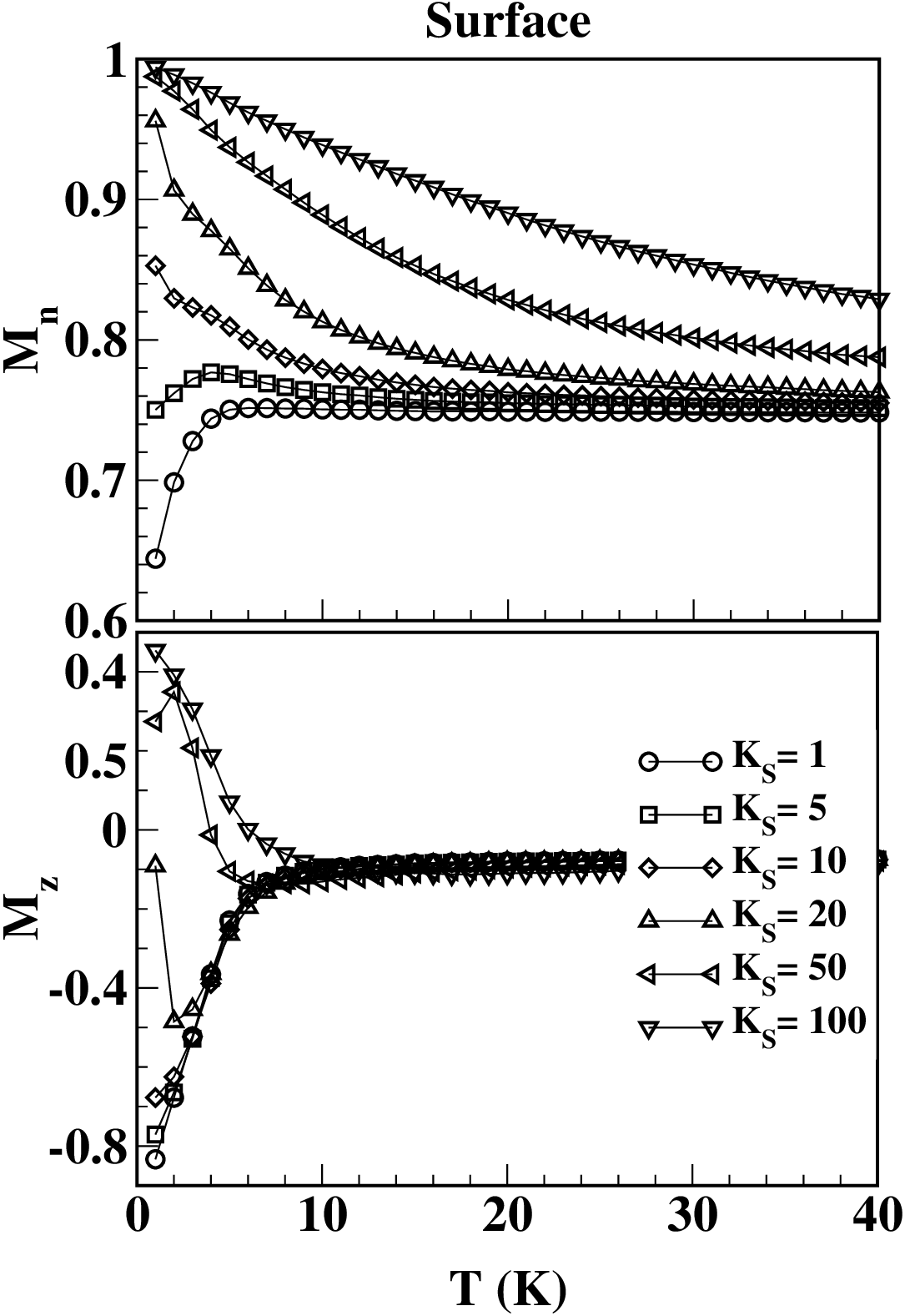}
\includegraphics[width=0.45\textwidth]{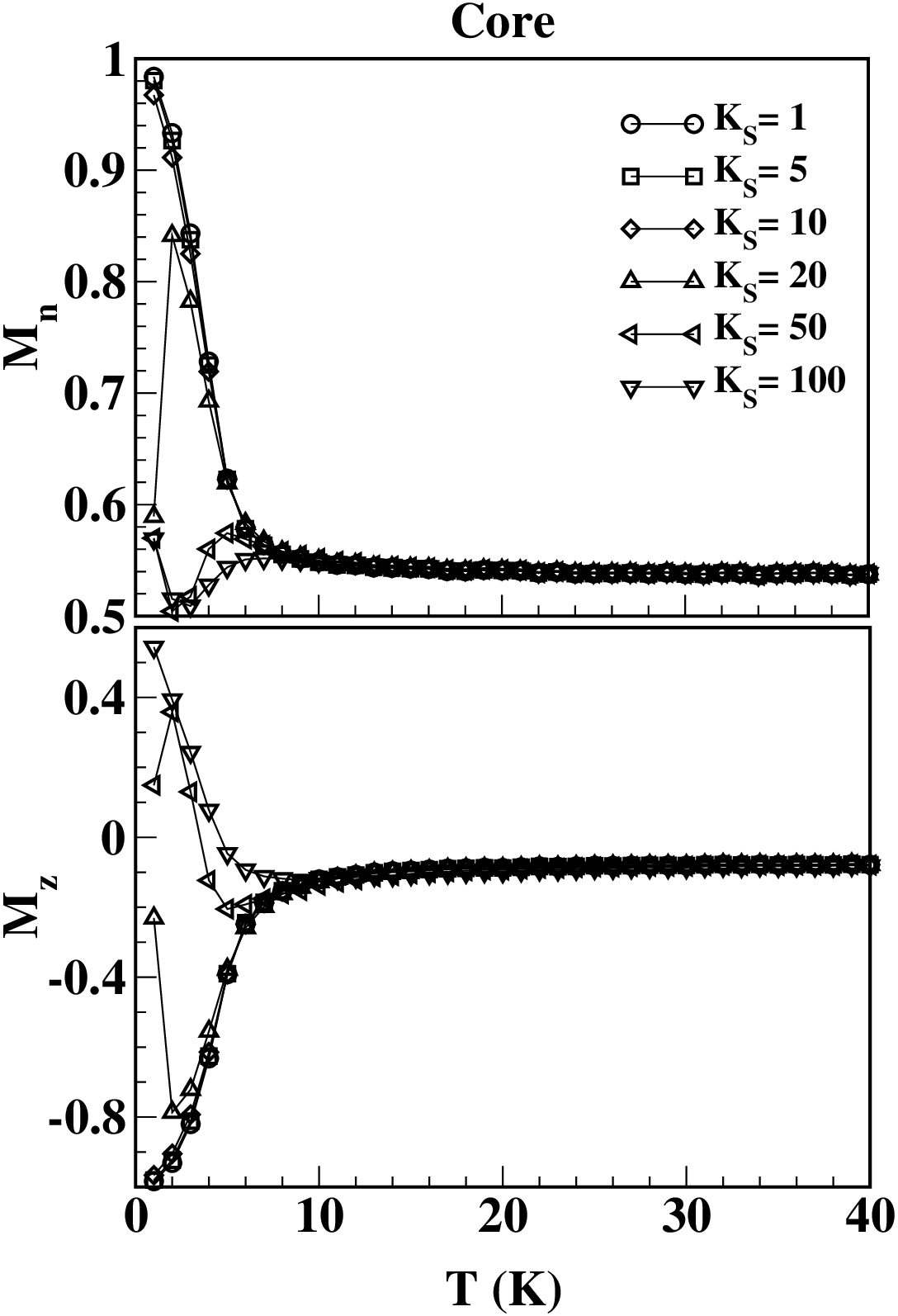}
\caption{ 
}
\label{EQ_FM_fig}
\end{figure}

For a better understanding of the role of surface anisotropy on the zero temperature configurations, we will first show the results of simulations for a spherical particle with $D=3$ and $k_C=1$, and the same structure than maghemite but with equal ferromagnetic (FM) interactions ($J_{ij}= 1$) among the spins instead of the real values. In Fig. \ref{EQ_FM_fig} we show the surface and core contributions to thermal dependence of the magnetization during the above-mentioned simulated annealing procedure for a range of values of $k_S$ spanning two orders of magnitude. 
The increase of $M_z$ as $T$ is reduced indicates that ordered states with FM order start to appear below $T\sim 10$ K which finally converge towards the equilibrium state at zero temperature . As reflected by the $M_z$ and $M_n$ values attained at $T=0$, the increase in $k_S$ results in a change in the kind of low $T$ order. For $k_S\lesssim 10$, quasi-uniform magnetization configurations are stable with the core spins pointing along the z axis ($M_n\lesssim 1$ and $|M_z| \sim 1$ as $T$ approaches zero in the right column curves of Fig. \ref{EQ_FM_fig}) and groups of surface spins slightly deviated towards the radial direction ($M_n\gtrsim 0.5$ and $|M_z| \sim 1$ as $T$ approaches zero in the left column curves of Fig. \ref{EQ_FM_fig}). However, as $k_S$ is further increased from $10$, hedgehog-like structures of surface spins start to form (note that $M_n$ at $T=0$ tends to $1$ at the surface) inducing the deviation of core spins towards the radial direction (indicated by the approach of $M_n$ to $0.5$ and of $M_z$ to $0$ as $T$ decreases for core spins). Snapshots of configurations representative of low and high $k_S$ values are depicted in Fig. \ref{Configs_fig} (upper rows) in qualitative agreement with simulations performed by Labay et al. for a FM particle with sc lattice \cite{Labayjap02}.

A similar phenomenon occurs in the case of a ferrimagnetic maghemite spherical particle with the real antiferromagnetic (AF) exchange constants given in the preceding section. However now, due to the dominant AF intersublattice coupling, there is a tendency to achieve antiparallel alignment among spins in different sublattices that causes a greater degree of disorder at the surface with respect to the FM case. Notice also that, even for small $k_S$, $M_z$ does not approach $1$ as $T$ tends to zero as in the FM case, but instead tends to the value of magnetization of the noncompensated spins ($M_{Unc}=(N_O-N_T)/N_{Total}$), that for a particle of diameter $D=3$ is $M_{Unc}= 0.285 (0.412)$ for surface (core) spins \cite{Iglesiasprb01}.
Moreover, for the ferrimagnetic particle, there is a change in the magnetic order  
at low $T$ as the $k_S/k_C$ ratio increases.
When surface anisotropy is small, the particle orders into a quasi-AF state in which spins in each sublattice are almost aligned along the core easy-axis. When increasing $k_S$, surface spins start to form throttled structures as those seen for $k_S$ in Fig. \ref{Configs_fig} that depart the spins from the z-axis. Finally, for $k_S >> k_C$, hedgehog-like configurations are favoured by the dominant radial anisotropy contribution (see the configuration for $k_S=100$).

In this case, the direct visualization of equilibrium configurations presented in Fig. \ref{Configs_fig} shows that the reduction of the saturation magnetization with particle size observed experimentally in different fine particles of ferrimagnetic oxides, can be attributed to the random canting of surface spins caused by the competing antiferromagnetic interaction between sublattices \cite{Kodamaprl96,Jiangjpcm99}. Moreover, as the results of our simulations confirm, the degree of disorder at the surface is larger than for the FM particle due to the complex interplay between the AF intralattice interactions and the local anisotropy easy-axis.

In order to study the influence of surface anisortropy on the magnetization processes, we have also simulated hysteresis loops at zero temperature. The loops have been computed by starting from a saturated state achieved after application of a high enough field of $h= 200$ K along the z axis and decreasing the field in constant steps $\delta h= 1$K, during which the magnetization was averaged over $1000$ MC steps after thermalization. 
The results for a spherical particle with $D=3$ and FM interactions are shown in Fig. \ref{Loops_SphericalFM_fig}. 
For a FM particle, the hysteresis loops are dominated by the surface contribution for all the values of $k_S$ studied as indicated by the non-squaredness of the loops around the coercive field. For high values of the surface anisotropy ($k_S=50, 100$), a magnetic field as high as $h= 20$ K is able to saturate the core but the surface spins instead point along the radial direction during the magnetization process. 
This is more clearly reflected in the right panels of Fig. \ref{Loops_SphericalFM_fig}, where we see that for high $k_S$, $M_n$ remains close to $1$ at the surface during all the reversal process, while the core spins depart from their easy directions ($M_n \sim 1$) dragged by the surface towards the radial direction close to the coercive field, where $M_n\sim 0.5$.
The hysteresis loops for the FM particle are to be compared with those for the ferrimagnetic maghemite particle presented in Fig. \ref{Loops_Spherical_fig} for $k_S= 10, 100$.
First, although the recersal process is still dominated by the surface spins, notice that now the loops for the ferrimagnetic particle become more elongated and have higher closure fields, ressembling those observed experimentally \cite{Troncjm03}. 
The increase of $k_S$ at the surface results in an increase in the coercive field $h_C$ and a reduction in the high field susceptibility. These features are due to the dominance of surface anisotropy over exchange interactions that create surface disordered states which become more difficult to reverse by the magnetic field.
More importantly, there is a change in the magnetization reversal mechanism when increasing $k_S$ from $10$ to $100$. In the first case (left panels in Fig. \ref{Loops_Spherical_fig}), the core and surface have similar $h_C$ and closure fields. The particle core reverses in a quasi-uniform form with the spins pointing mostly along the z-axis ($M_n\approx 1$) except near $h_c$, and with the surface spins following the core reversal (with $M_n<1$ indicating alignment close to the z direction) \cite{WWW_HMM2003}. 
However, at higher $k_S$ (right panels in Fig. \ref{Loops_Spherical_fig}), surface spins remain close to the local radial direction ($M_n\approx 1$) during all the reversal process, driving the core spins out of their local easy axis and making their reversal non-uniform ($0.5<M_n<1$) due to the appearance of the hedgehog-like structures during the reversal \cite{WWW_HMM2003}.

In conclusion, we have studied the differences in the equilibrium states and reversal processes of ferrimagnetic with respect to FM particles, showing in both cases that finite-size effects, together with the existence of surface disorder due to increased anisotropy at the surface, are necessary ingredients for the low temperature anomalous magnetic properties observed experimentally in nanoparticle systems.  

\section*{Acknowledgements}
We acknowledge CESCA and CEPBA under coordination of 
C$^4$ for the computer facilities. This work has been supported by 
SEEUID through project MAT2000-0858 and CIRIT under project 2001SGR00066.






\newpage
\section*{FIGURE CAPTIONS}
FIG. 1: Spherical ($D=3$) particle with FM interactions ($J_{ij}=1$): thermal dependence of the surface (left column) and core (right column) contributions to the projection of the magnetization along the z axis $M_z$ and to $M_n$ as defined in the text during a progressive cooling from a high $T$ at a constant rate $\delta T= -1$ K.

FIG. 2: Spherical ($D=3$) particle with the real maghemite AF interactions quoted in the text: thermal dependence of the surface (left column) and core (right column) contributions to the projection of the magnetization along the z axis $M_z$ and to $M_n$ as defined in the text during a progressive cooling from a high $T$ at a constant rate $\delta T= -1$ K.

FIG. 3: Snapshots of the equilibrium configurations attained after cooling from a high temperature disordered state for FM (upper row) and ferrimagnetic maghemite (lower row) spherical particles with $D=3$ and maghemite lattice structure. 

FIG. 4: Hysteresis loops for a spherical particle of diameter $D=3$ with FM interactions.
The core anisotropy constant is $k_C=1$, the results for several values of the surface anistropy constant are displayed $k_S= 1,5,10,20,50,100$. The magnetization component along the applied field direction is shown in lower right panel, the core contribution is displayed in the upper right panel. Right panels show the suface and core contributions to $M_n$ as defined in the text.

FIG. 5: Hysteresis loops for a spherical particle of diameter $D=3$ with the real values of exchange constants of maghemite. The main panels show the total magnetization along the field direction and the corresponding surface (dashed lines) and core (circles) contributions. The upper subpanels show the surface (circles) and core (continuous lines) contributions to $M_n$.
Left panel: $k_S=10$, right panel: $k_S=100$.

\newpage
\ 
\newpage
\ 
\begin{figure}[h] 
\centering 
\includegraphics[width=0.45\textwidth]{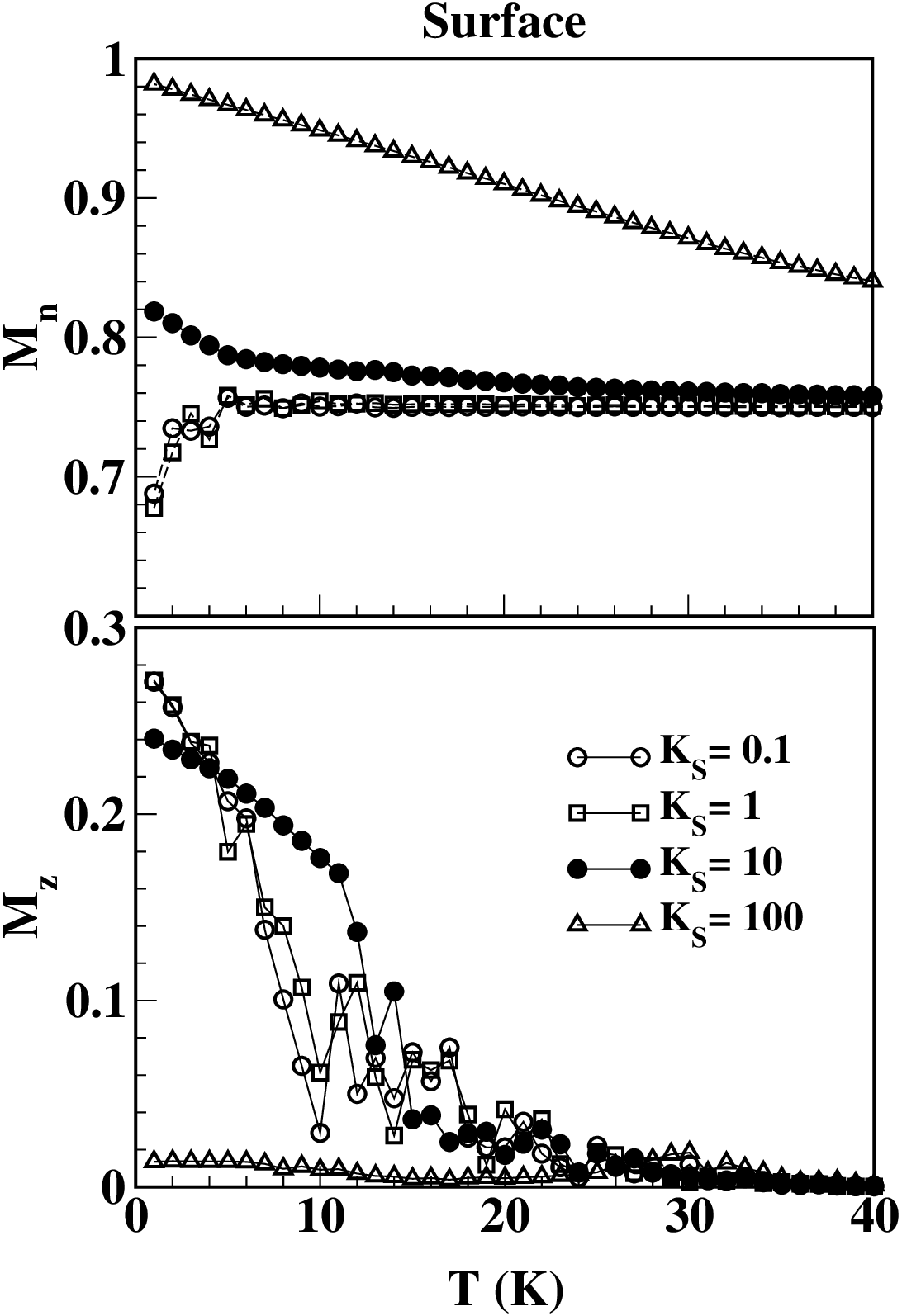}
\includegraphics[width=0.45\textwidth]{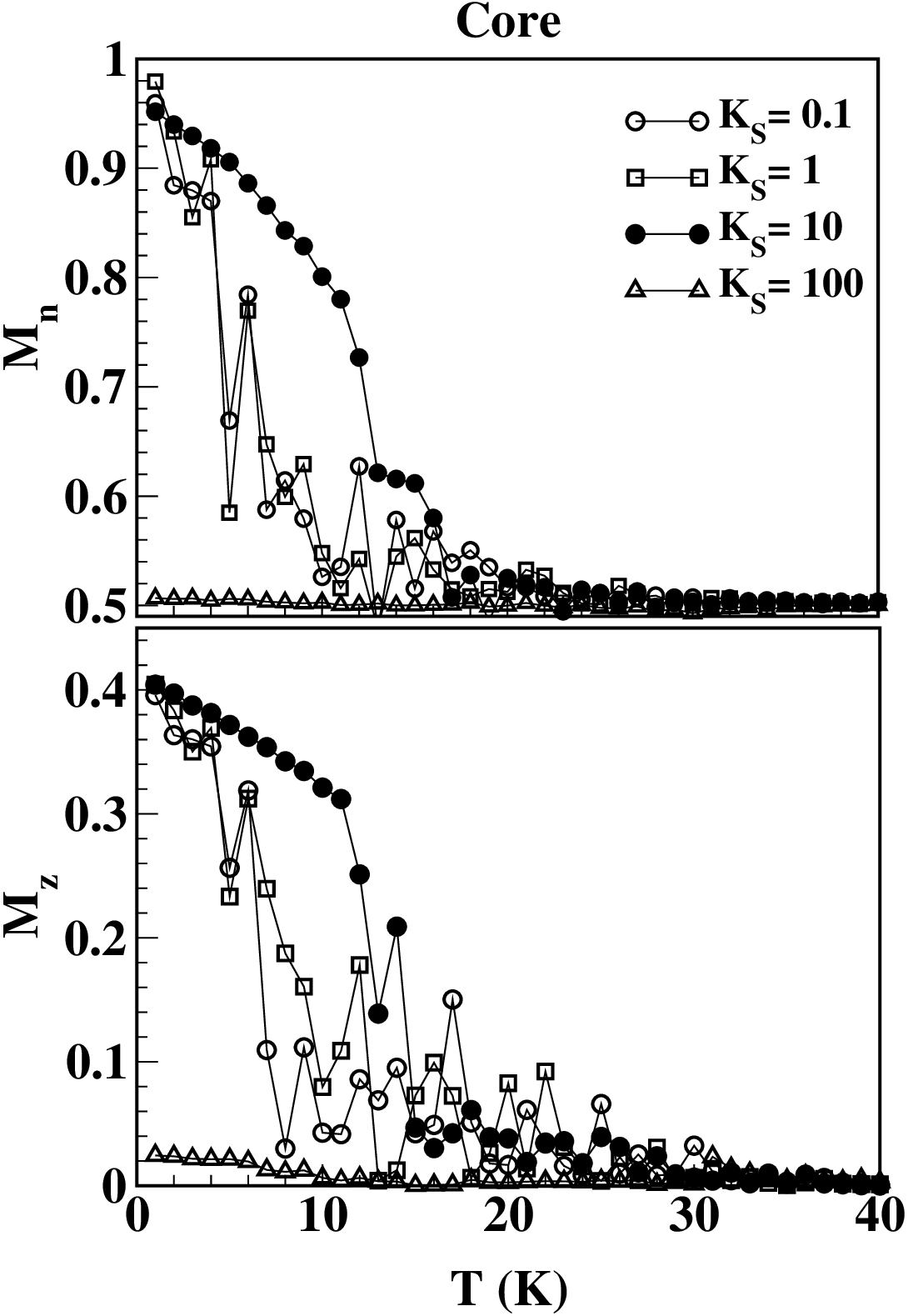}
\caption{ 
}
\label{EQ_Maghemite_fig}
\end{figure}
\newpage
\ 
\begin{figure}[h] 
\centering 
\caption{ 
}
\label{Configs_fig}
\end{figure}
\newpage
\ 
\begin{figure}[h] 
\centering 
\includegraphics[width=1.0\textwidth]{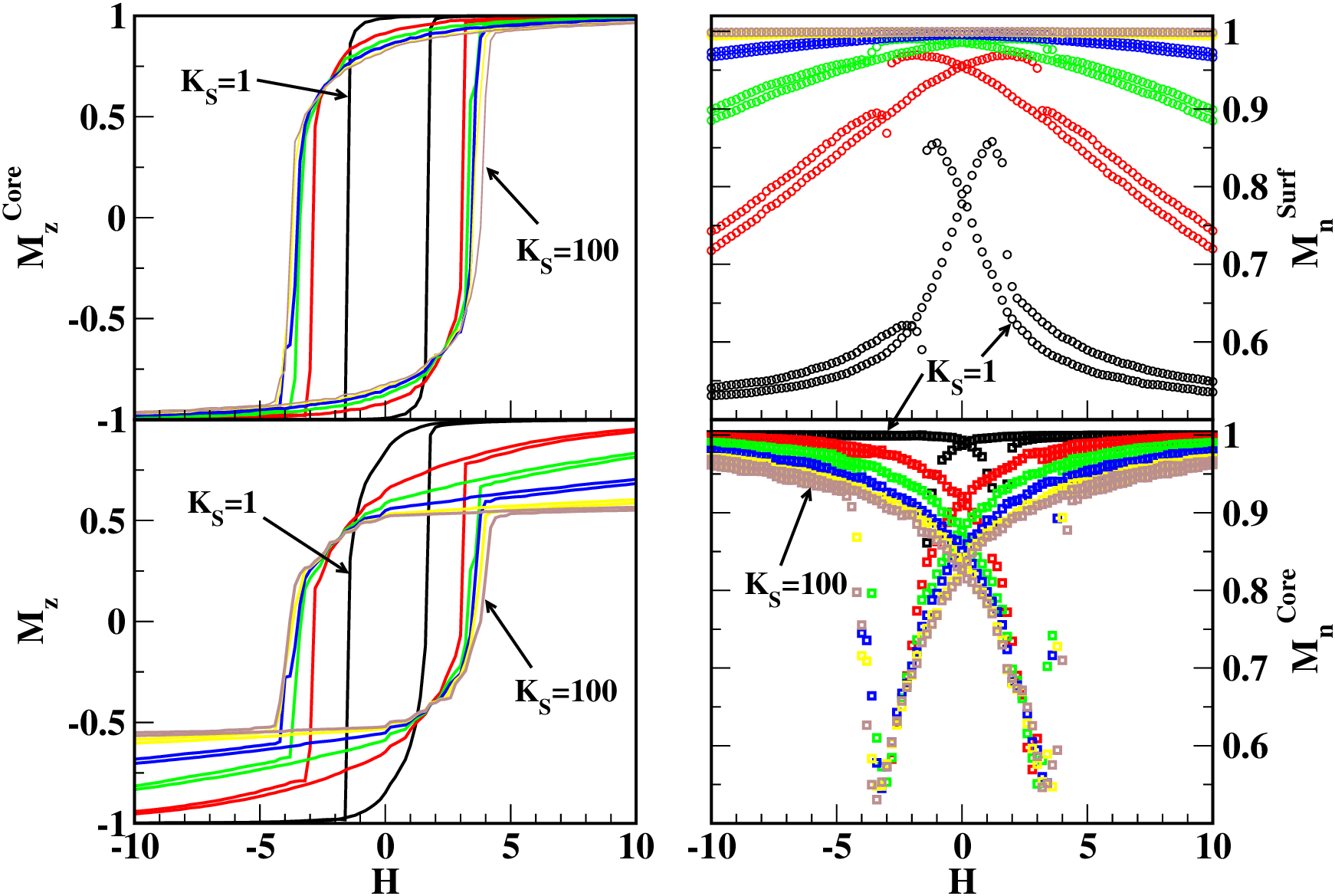}
\caption{ 
}
\label{Loops_SphericalFM_fig}
\end{figure}
\newpage
\
\begin{figure}[h] 
\centering 
\includegraphics[width=0.49\textwidth]{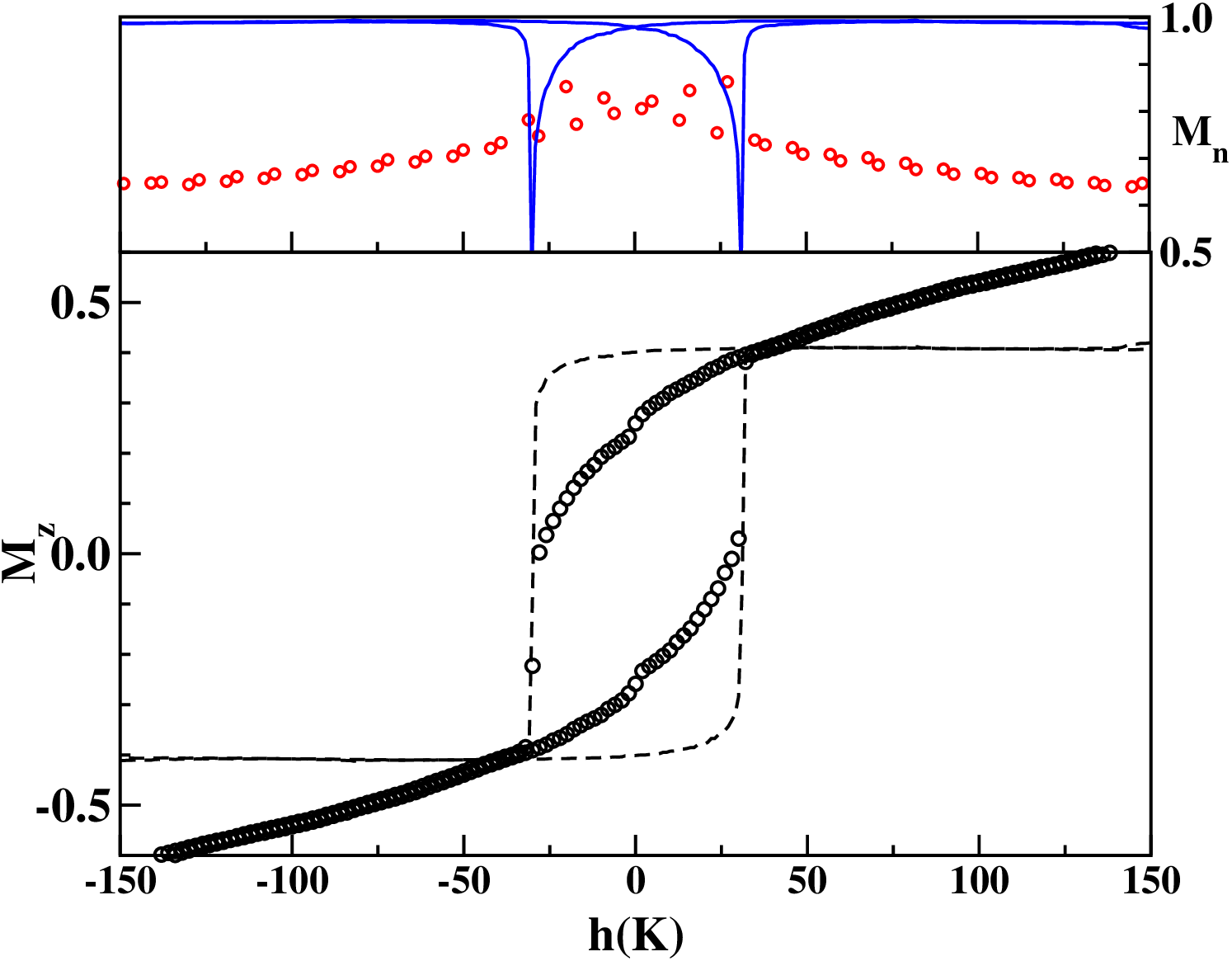}
\includegraphics[width=0.49\textwidth]{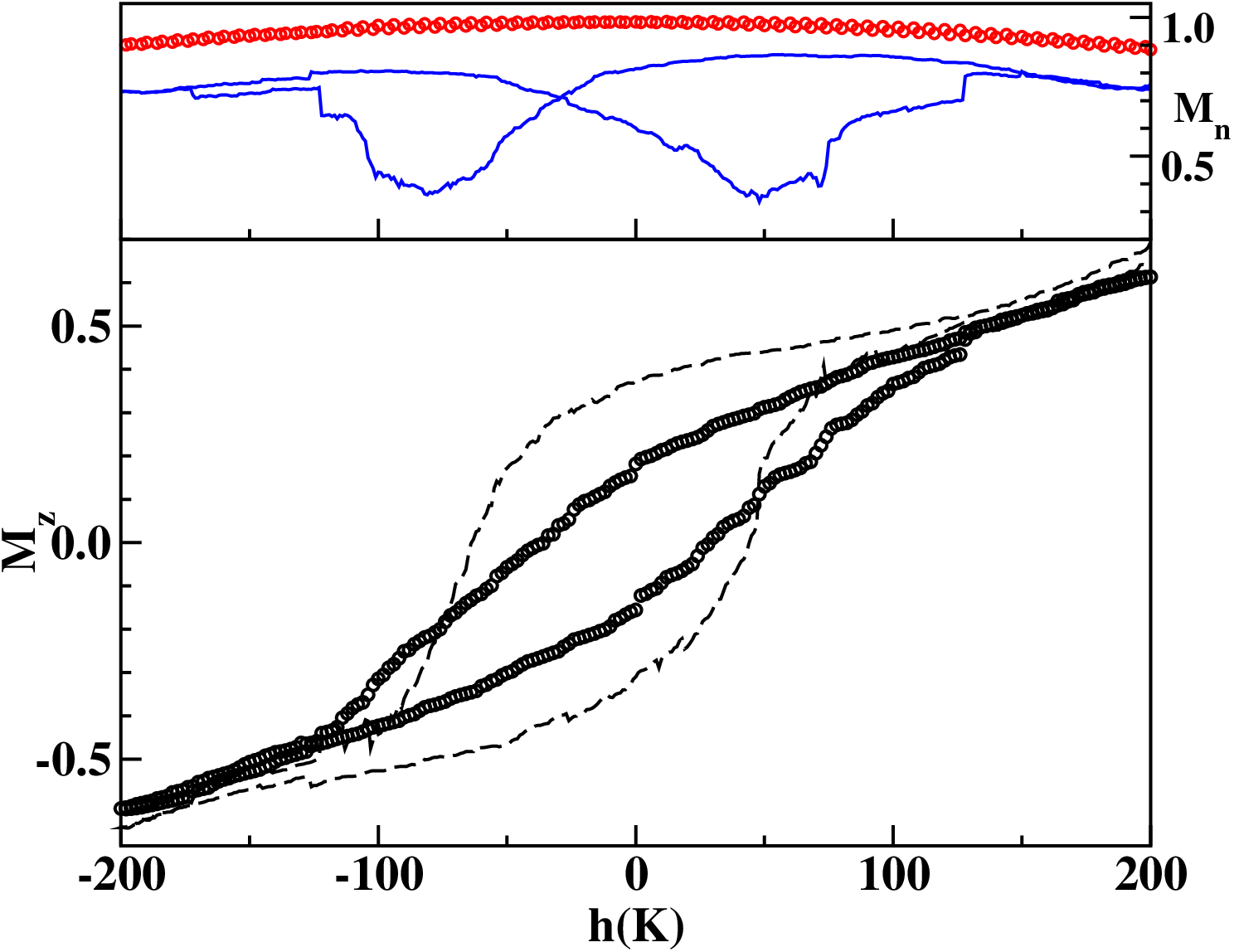}
\caption{ 
}
\label{Loops_Spherical_fig}
\end{figure}
\end{document}